\documentclass[conference]{IEEEtran}
\IEEEoverridecommandlockouts
\usepackage{cite}
\usepackage{amsmath,amssymb,amsfonts}
\usepackage{graphicx}
\usepackage{textcomp}
\usepackage{xcolor}
\usepackage{booktabs}
\usepackage{multirow}
\usepackage{url}
\usepackage{subfigure}

\usepackage{algorithm}
\usepackage{algorithm}
\usepackage{algorithmicx}
\usepackage{algpseudocode}
\usepackage{amsmath}

\usepackage{graphicx}

\def\BibTeX{{\rm B\kern-.05em{\sc i\kern-.025em b}\kern-.08em
    T\kern-.1667em\lower.7ex\hbox{E}\kern-.125emX}}
    
\begin{document}

\title{ Long-distance Deterministic Transmission among TSN Networks: Converging CQF and DIP\\
}

\author{\IEEEauthorblockN{Weiqian Tan\IEEEauthorrefmark{1},
Binwei Wu\IEEEauthorrefmark{2}
}
\IEEEauthorblockA{\IEEEauthorrefmark{1}Southeast University, Nanjing, China\\
\IEEEauthorrefmark{2}Purple Mountain Laboratories, Nanjing, China\\
Email: t\_w\_q@foxmail.com,
wubinwei163@163.com}}

\maketitle

\begin{abstract}
With the development of 5G, innovative applications requiring bounded transmission delays and zero packet loss emerge, e.g., AR, industrial automation, and smart grid. In this circumstance, time-sensitive networking (TSN) is proposed, which addresses the deterministic transmission in the local area networks. Nevertheless, TSN is essentially a Layer 2 technique, which cannot provide deterministic transmission on a large geographic area. To solve this problem, this paper proposes a hierarchical network for the end-to-end deterministic transmission. In the proposed network, we leverage CQF (i.e., one of the most efficient TSN mechanisms) in the access networks which aggregates the traffic from end-devices. Meanwhile, in the core network, we exploit the DIP (i.e., a well-known deterministic networking mechanism for backbone networks) for long-distance deterministic transmission.
We design the cycle alignment mechanism to enable seamless and deterministic transmission among hierarchical networks. A joint schedule is also formulated, which introduces the traffic shaping at the network edge to maximize the network throughput.
Experimental simulations show that the proposed network can achieve end-to-end deterministic transmission, even in the highly-load scenarios.

\end{abstract}

\begin{IEEEkeywords}
deterministic networking, long-distance end-to-end transmission, DIP, CQF
\end{IEEEkeywords}

\section{Introduction}

With the development of the Internet and communication networks, many time-sensitive applications are emerging, such as factory automation, connected vehicles and smart grids \cite{grossman2019deterministic}. Traditional IP networks performing best-effort transmission provide some Quality of Service (QoS) strategies (e.g., DiffServ and congestion control \cite{chettri2019comprehensive}). However, due to micro-burst traffic \cite{shan2019observing} existing in networks, these mechanisms can not achieve deterministic transmission, which means zero packet loss, bounded delays and jitters (i.e., delay variation) \cite{charny2000delay}. Thus, for eliminating micro-burst traffic, IEEE aims to develop a set of standards named Time-Sensitive Networking (TSN) to provide deterministic transmission in layer 2. Due to the high efficiency, Cyclic Queuing Forwarding (CQF) \cite{ieee8021qch} is one of the top choices in TSN \cite{nasrallah2019tsn}. Besides, Credit-Based Shaper \cite{ieee8021qav} and Time-Aware Shaper \cite{ieee8021qbv} are also well-known mechanisms in TSN. However, TSN cannot provide long-distance deterministic transmission because of the requests for high precision time synchronization and low propagation delays, which leads to the ``information isolated island" among TSN networks.
Global management of TSN applications (e.g. teleprotection, wind farm applications, etc. \cite{salazar2019white}) requests long-distance deterministic transmission among TSN networks.

For achieving large-scale deterministic transmission, the IETF Deterministic Networking (DetNet) group focus on promoting the standardization in layer 3. The standards of DetNet are still under discussion. Some drafts have been proposed, such as Cycle Specified Queuing and Forwarding (CSQF) \cite{chen2018segment} and Deterministic IP (DIP) \cite{Qiang2019LDNforwarding}. CSQF leverages Segment Routing \cite{SegmentRoutingArchi} Identifier to coordinate cyclic transmission times across large-scale networks to offer deterministic transmission. DIP is one of the effective mechanisms in DetNet. The routers performing DIP (named as DIP routers) achieve frequency synchronization. Packets are transported in a slotted fashion by carrying a cycle identifier. After investigating recently proposed deterministic forwarding mechanisms, we find that no researches have been conducted to apply DetNet to achieve long-distance deterministic transmission among TSN networks. In this paper, we propose a hierarchical network for long-distance end-to-end deterministic transmission. CQF technique is used in the access layer, and DIP supports the core layer.

Due to the differences between CQF and DIP, it is not effective to integrate these two mechanisms directly. The first problem is different time synchronization mechanisms in CQF and DIP. Time synchronization with high precision used in most TSN mechanisms is hard to achieve in long-distance deterministic transmission.
Thus, DIP only requests frequency synchronization. The second problem is the duration of a cycle may be different in CQF and DIP. These two problems are the barriers to construct correspondences between cycles in CQF switches and DIP routers because cycles in CQF do not correspond to cycles in DIP one by one. However, cycle correspondences are the foundation to achieve deterministic transmission in CQF and DIP.

To tackle the problems mentioned above, this paper establishes the network model of the hierarchical network and proposes a cycle alignment mechanism to establish the correspondences between cycles in CQF and DIP. Based on the cycle alignment, we propose a traffic shaping mechanism to solve the disorder of traffic at the edge of access and core networks to achieve end-to-end joint scheduling across CQF and DIP. The joint scheduling is formulated as integer linear programming to maximize the number of acceptable time-sensitive flows. In the simulations, the scheduling is conducted in a greedy algorithm. The time-sensitive flows (TS flows) with high weight are scheduled preferentially. For each TS flow, the scheduling chooses the transmission pattern assuming minimum resources.

The rest of this paper is organized as follows. Section \ref{section2} details the deterministic networking technologies used in this paper (i.e., CQF and DIP). In Section \ref{section3}, the network model of the hierarchical network is introduced. Section \ref{section4} proposes the end-to-end joint scheduling approach. Section \ref{section5} shows the setup and the results of simulations. Finally, Section \ref{section6} concludes this paper.

\section{Background}\label{section2}

\subsection{Cyclic Queuing Forwarding}
IEEE 802.1Qch (CQF) standard \cite{ieee8021qch} proposes a method to synchronize enqueue and dequeue operations, and schedule the transmission in a cyclic fashion. Time in CQF networks is divided into cycles with equal duration $d$. All CQF switches in the same CQF network are perfectly synchronized. There exist two queues ($q1$ and $q2$) in a CQF switch output port. In cycle $c$, queue $q1$ is open for dequeuing packets. At the same time, queue $q2$ enqueues packets from upstream. In the next cycle $c+1$, queue $q1$ enqueues packets and queue $q2$ dequeues packets. Due to the cyclic manner for transmission, the packets received in cycle $c$ are scheduled to be retransmitted in cycle $c+1$, and the maximum possible delay experienced in a CQF switch is from the beginning of cycle $c$ to the end of $c+1$, equal to $2d$.

The upper and lower bounds are easily calculated in CQF networks.
The number of hops is defined as $h$ 
. $D_{\rm{max}}$ and $D_{\rm{min}}$ represent the maximum and minimum latency respectively. The equations to calculate $D_{\rm{max}}$ and $D_{\rm{min}}$ are as follows:
\begin{equation}\label{equ::Dmax}
    D_{\rm{max}} = (h + 1) \times d
\end{equation}
\begin{equation}\label{equ::Dmax}
    D_{\rm{min}} = (h - 1) \times d
\end{equation}


\subsection{Deterministic IP}

\begin{figure}[htbp]
\centerline{\includegraphics[width=.65\linewidth]{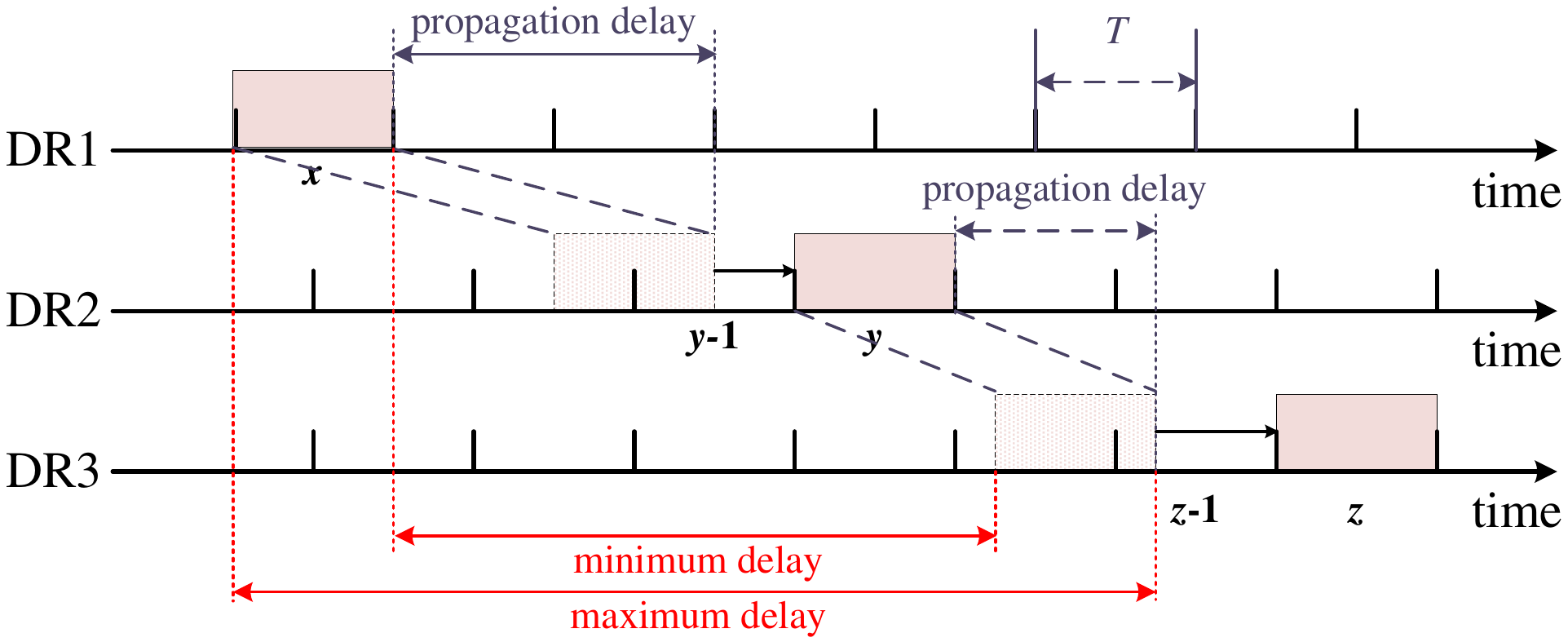}}
\caption{An example of transmission among DIP routers. The packets sent in cycle $x$ by DR1 will be retransmitted in cycle $y$ by DR2. The transmission cycle in DR3 is $z$. }
\label{fig::dipIntro}
\end{figure}

Time in DIP is also divided into cycles with equal duration $T$. Because perfect time synchronization is hard to achieve in large geographical coverage, DIP only requests frequency synchronization and permits the propagation delay exceeding the cycle duration. DIP leverages correspondences between cycles to achieve deterministic transmission. Fig.~\ref{fig::dipIntro} shows the DIP transmission among three DIP routers (DR1, DR2, and DR3). Based on the propagation delays between DR1 and DR2, and the one between DR2 and DR3, the following cycle correspondences are constructed: the cycle $x$ in DR1 is mapped to cycle $y$ in DR2, and the cycle $y$ is mapped to cycle $z$ in DR3. These cycle correspondences imply that the packets sent in cycle $x$ will be received in DR2 not later than cycle $y-1$, and retransmitted to DR3 in cycle $y$ in DR2. It is similar between DR2 and DR3.

The packets sent at cycle $x$ from DR1  carry the cycle identifier $x$. After receiving them, DR2 will check the cycles correspondences table stored in it, and find the correspondence $x \rightarrow y$. Then DR2 enqueues these packets to the queue corresponding to cycle $y$. At cycle $y$, the queue will open and dequeue these packets to DR3. Fig.~\ref{fig::dipIntro} shows the maximum and minimum delay from DR1 to DR3, and the maximum jitter is less than $2T$.

\section{Hierarchical Network}\label{section3}
\begin{figure}[h]
\centering
\includegraphics[width=.75\linewidth]{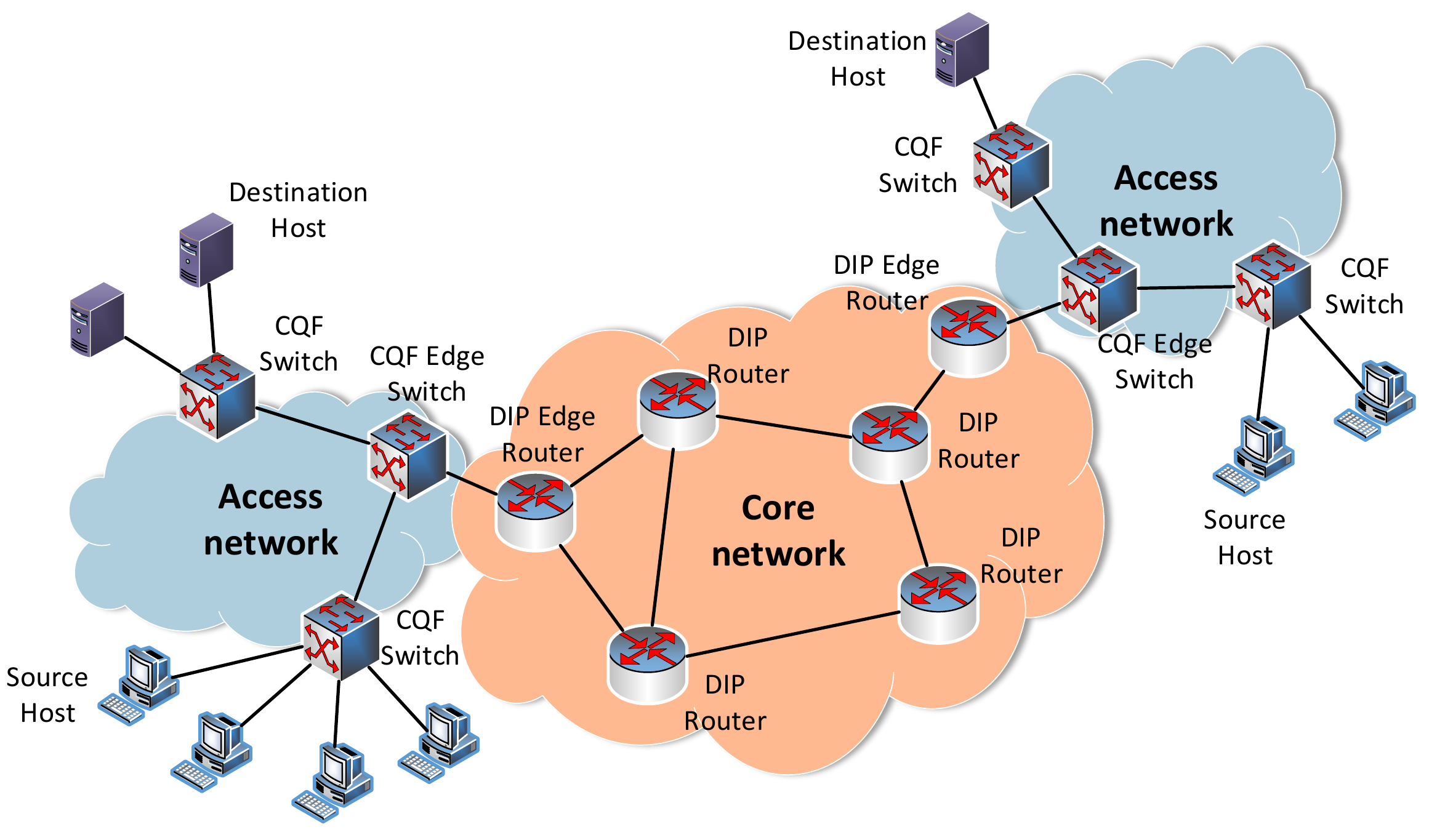}
\caption{
The hierarchical network. In the access networks, CQF guarantees deterministic transmission in the local area. In the core network, DIP provides long-distance deterministic transmission. Hosts and CQF (edge) switches perform CQF, while DIP (edge) routers perform DIP. Especially, CQF edge switches and DIP edge routers perform traffic shaping and the cycle alignment across CQF and DIP. 
}
\label{fig::structureOfConvergedNetworks}
\end{figure}

We propose a hierarchical network to achieve long-distance deterministic transmission among TSN networks, as shown in Fig.~\ref{fig::structureOfConvergedNetworks}. The access networks perform CQF and the core network forwards packets following DIP. The following subsections will demonstrate the transmission mechanisms in the hierarchical network systematically.

\subsection{Network model}\label{section3.1}

The whole network is represented as $G=\{V,E\}$. $V$ is the set of network devices, and $E$ is the set of links. Network devices contain source hosts, destination hosts, CQF switches, CQF edge switches, DIP routers, and DIP edge routers, i.e., $V=\{V_{\rm{src}}, V_{\rm{dest}}, V_{\rm{cqf}}, V_{\rm{cqf}}^{\rm{edge}}, V_{\rm{dip}}, V_{\rm{dip}}^{\rm{edge}}\}$. The processing delays in all devices are ignored. 

For a link $e \in E$, source $v_1$ and end $v_2$ uniquely identify the link, i.e., $e=(v_1, v_2)$. A link $e$ corresponds to a propagation delay $\Delta_e$. 
The bandwidth of link $e$ is denoted as $BW_e$. 

A path from the source $v_0 \in V_{\rm{src}}$ to the destination $v_{|p|} \in V_{\rm{dest}}$ is denoted by $p=(v_0, v_1, \cdots, v_{|p|})$, where $|p|-1$ is the number of intermediate nodes along path $p$. A set $P$ contains all possible paths from $V_{\rm{src}}$ to $V_{\rm{dest}}$, i.e., $p \in P$.

Time-sensitive applications located in source hosts generate time-sensitive flows (TS flows). $F$ represents the set of TS flows. Each TS flow $f \in F$ is periodic, and the period is $\Delta_f$. A TS flow $f$ emits packets in size of $\omega_f$ bits during every $\Delta_f$. Moreover, the source and destination of a flow $f$ are deterministic.
The maximum acceptable end-to-end delay of $f$ is $\Delta_{f}^{\rm{e2e}}$.

\subsection{Time models in CQF and DIP}\label{section3.2}    

\begin{figure}[htbp]
\centerline{\includegraphics[width=.80\linewidth]{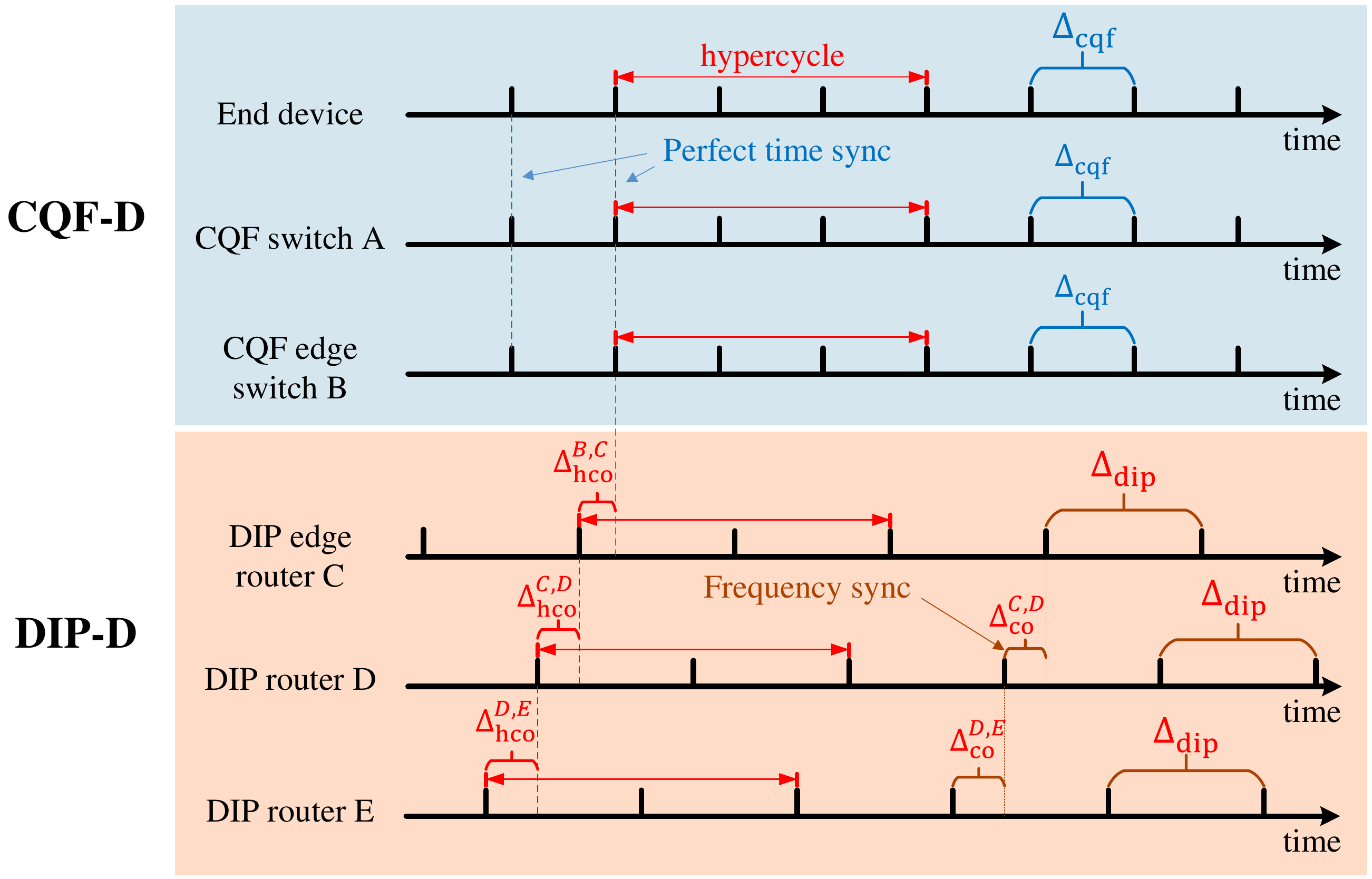}}
\caption{Time models in the hierarchical network. In CQF-D, all devices achieve perfect time synchronization, while DIP-D only requires frequency synchronization. $N_{\rm{cqf}} = 3$ and $N_{\rm{dip}} = 2$ in this figure. 
}
\label{fig::timeModels}
\end{figure}

Time models in CQF and DIP are different. We define every access and core network as a time-domain named CQF-D and DIP-D respectively, as shown in Fig.~\ref{fig::timeModels}. For uniformly scheduling, we introduce \textit{hypercycle}, which contains multiple cycles in CQF and DIP.

{\bf{CQF-D}:} CQF-D contains source/destination hosts, CQF edge switches and CQF switches. Fig.~\ref{fig::timeModels} shows that, in CQF-D, time is divided into time cycles with the same duration $\Delta_{\rm{cqf}}$. A hypercycle contains $N_{\rm{cqf}}$ consecutive cycles. The length of a hypercycle is $\Delta_{\rm{hc}} = N_{\rm{cqf}}\Delta_{\rm{cqf}}$. All devices in CQF-D achieve perfect time synchronization, which means that the start time of a cycle (also a hypercycle) in all devices is the same.

{\bf{DIP-D}:} DIP-D covers DIP routers and DIP edge routers. Similar to CQF-D, time in DIP-D is divided into time cycles with the same duration $\Delta_{\rm{dip}}$. A hypercycle in DIP-D contains $N_{\rm{dip}}$ consecutive DIP cycles. Thus, the length of a hypercycle is $\Delta_{\rm{hc}} = N_{\rm{dip}}\Delta_{\rm{dip}}$. As shown in Fig.~\ref{fig::timeModels}, DIP-D only requires frequency synchronization. For a pair of adjacent devices $(A,B)$ in DIP-D, there may exist a constant gap between the start time of cycles in $A$ and $B$. The gap is defined as the offset of cycles, $\Delta_{\rm{co}}^{A,B}$. $\Delta_{\rm{co}}^{A,B}$ is the difference between the start time of cycles in $A$ and $B$, and $\Delta_{\rm{co}}^{A,B} \geq 0$.  


{\bf{Hypercycle}:} For allocating deterministic cycles to each TS flow, the length of a hypercycle should satisfy:
\begin{equation}\label{equ::durationHC}
\Delta_{\rm{hc}} = N_{\rm{cqf}}\Delta_{\rm{cqf}} = N_{\rm{dip}}\Delta{\rm{dip}} = N_F\Delta_f
\end{equation}
where $N_{\rm{cqf}}, N_{\rm{dip}}, N_F \in \mathbb{Z}^+$, $N_{\rm{cqf}}$ and $N_{\rm{dip}}$ are greater than the number of queues in output ports of CQF switches and DIP routers respectively.


Fig.~\ref{fig::timeModels} illustrates hypercycles in CQF-D and DIP-D. The offset of hypercycle between a pair of adjacent devices $(A,B)$ is defined as $\Delta_{\rm{hco}}^{A,B}$. $\Delta_{\rm{hco}}^{A,B}$ is the difference between the start time of hypercycles in $A$ and $B$, and $\Delta_{\rm{hco}}^{A,B} \geq 0$.

            

\subsection{Cycle alignment}\label{section3.3}
We design a cycle alignment mechanism to construct correspondences between cycles in different devices. The cycle alignment is defined as follows: if packets sent in cycle $a$ in a node $A$ will all be received no later than cycle $b$ in the downstream node $B$, we define a cycle alignment relationship as $\Phi_{(A,B)}(a)=b$.

{\bf{Alignment in CQF-D}:} Inside CQF-D, packets transmitted by a node during cycle $i$ must be received by the downstream node during the same cycle and retransmitted in cycle $i+1$.
Thus, the cycle alignment in CQF-D is $\Phi(i)=i$.

\begin{figure}[htbp]
\centerline{\includegraphics[width=.75\linewidth]{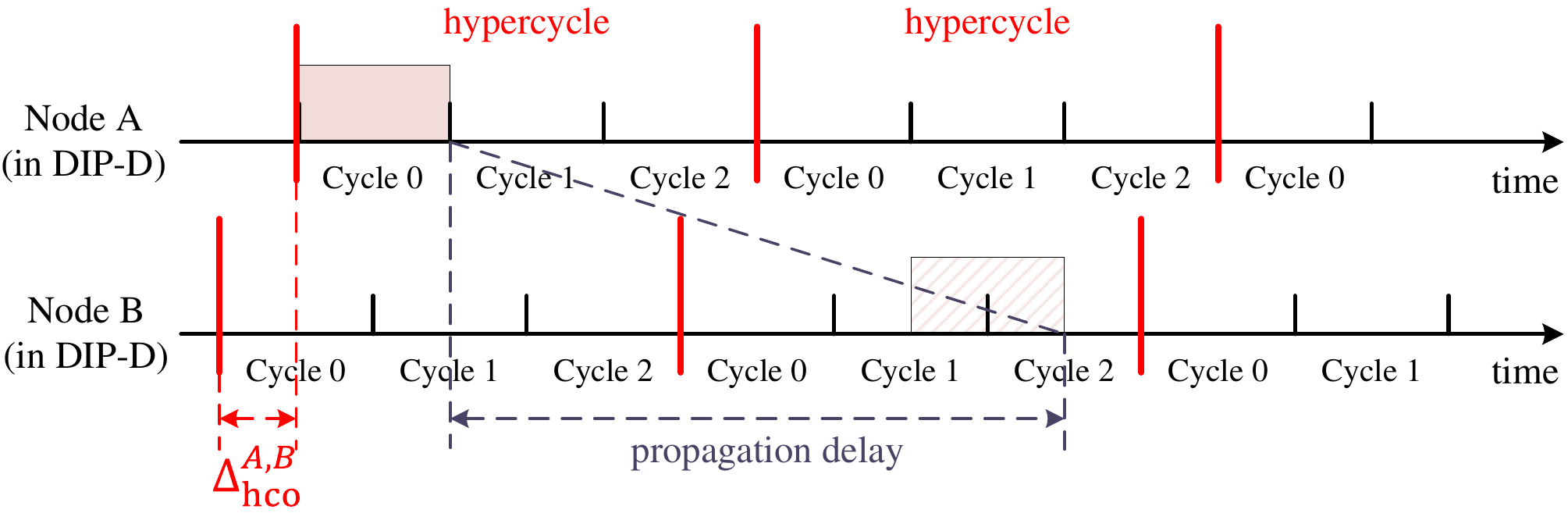}}
\caption{The cycle alignment in DIP-D. All packets sent in node $A$'s cycle 0 will be received no later than node $B$'s cycle 2. The cycle alignment $\Phi_{(A,B)}(0) = 2$ is established.   }
\label{fig::cycleMappingDIP}
\end{figure}

{\bf{Alignment in DIP-D}:} Fig.~\ref{fig::cycleMappingDIP} illustrates a cycle alignment in DIP-D. There exist adjacent DIP devices node $A$ and $B$ with the same cycle length $\Delta_{\rm{dip}}$. 
The length of a hypercycle is $\Delta_{\rm{hc}} = 3\Delta_{\rm{dip}}$ (i.e., $N_{\rm{dip}} = 3$). Cycle 0 in node $A$ is mapped to cycle 2 in node $B$ (i.e., $\Phi_{(A,B)}(0) = 2$).

The cycle alignment $\Phi_{(A,B)}(0) = 1$ can be calculated based on the duration of cycles in DIP-D $\Delta_{\rm{dip}}$, the length of hypercycle $\Delta_{\rm{hc}}$, the offset of hypercycles $\Delta_{\rm{hco}}^{A,B}$, and the propagation delay $ \Delta_{(A,B)}$. $\Phi_{(A,B)}(\cdot)$ is defined as
\begin{equation}
    \Phi_{(A,B)}(x) = \lceil \varphi_{(A,B)}(x) - 1 \rceil \, {\rm{mod}} \, N_{\rm{dip}}
\end{equation}
where
\begin{equation}
\resizebox{.9\hsize}{!}{$\varphi_{(A,B)}(x)=\frac{((x+1) \, {\rm{mod}} \, N_{\rm{dip}})\cdot \Delta_{\rm{dip}} + \Delta_{(A,B)}+\Delta_{\rm{hco}}^{A,B}}{\Delta_{\rm{dip}}}$}
\end{equation}
Note that the index $x$ loops from 0 to $N_{\rm{dip}}-1$.

\begin{figure}[htbp]
\centerline{\includegraphics[width=.75\linewidth]{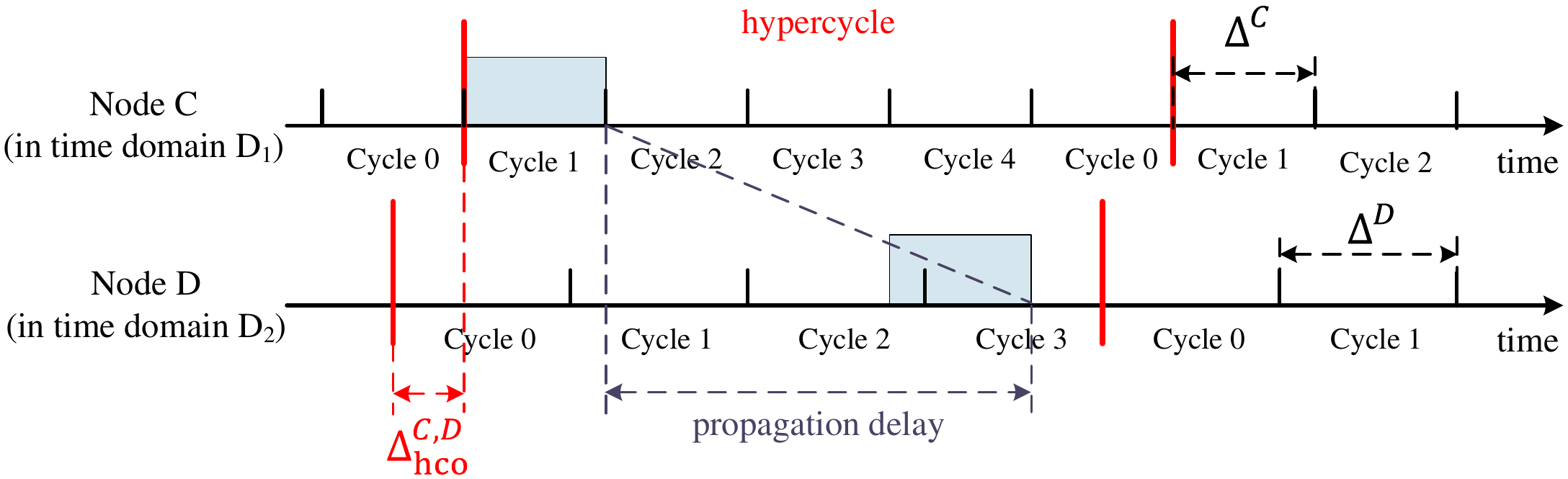}}
\caption{Cycle alignment across CQF-D and DIP-D. Nodes $C$ and $D$ are in different time domains. Cycle 1 in node $C$ is mapped to cycle 3 in node $D$.  }
\label{fig::cycleMappingEdge}
\end{figure}

{\bf{Alignment across CQF-D \& DIP-D}:}  Fig.~\ref{fig::cycleMappingEdge} illustrates an established cycle alignment from node $C$ to its adjacent node $D$. Node $C$ and $D$ are located in different time domains. The length of cycles in node $C$ and $D$ are $\Delta^{C}$ and $\Delta^{D}$ respectively. The length of hypercycle is $\Delta_{\rm{hc}}=5\Delta^{C}=4\Delta^{D}$, and the offset of hypercycle is $\Delta_{\rm{hco}}^{C,D}$. Cycle 0 in node $C$ is mapped to cycle 3 in node $D$, i.e., $\Phi_{(C,D)}(1)=3$.

Due to the inequality of cycle durations in CQF-D and DIP-D, $\Delta_{\rm{co}}^{C,D}$ is not constant. However, the offset of hypercycles between CQF-D and DIP-D is a constant. Taking Fig.~\ref{fig::cycleMappingEdge} as an example, the cycle alignment can be calculated based on $\Delta_{\rm{hco}}^{C,D}$, $\Delta^{C}$, $\Delta^{D}$, $\Delta_{\rm{hc}}$, and $\Delta_{(C,D)}$. The cycle alignment across CQF-D and DIP-D, $\Phi_{(C,D)}(\cdot)$, is defined as
\begin{equation}
    \Phi_{(C,D)}(x) = \lceil \varphi_{(C,D)}(x) - 1 \rceil \, {\rm{mod}} \, \frac{\Delta_{\rm{hc}}}{\Delta^{D}}
\end{equation}
where
\begin{equation}
\resizebox{.9\hsize}{!}{$\varphi_{(C,D)}(x)=\frac{((x+1) \, {\rm{mod}} \, \frac{\Delta_{\rm{hc}}}{\Delta^{C}})\cdot \Delta^{C} + \Delta_{(C,D)}+\Delta_{\rm{hco}}^{C,D}}{\Delta^{D}}$}
\end{equation}
Note that the index $x$ loops from 0.


\subsection{Traffic shaping}\label{section3.4}
The hierarchical network leverages cycle shifting to achieve traffic shaping. A cycle shift represents the additional cycles for retransmission based on the result of $\Phi(\cdot)$. Cycle shifts in sources, CQF switches/DIP routers, and CQF edge switches/DIP edge routers are different respectively.

\begin{figure}[htbp]
\centerline{\includegraphics[width=.75\linewidth]{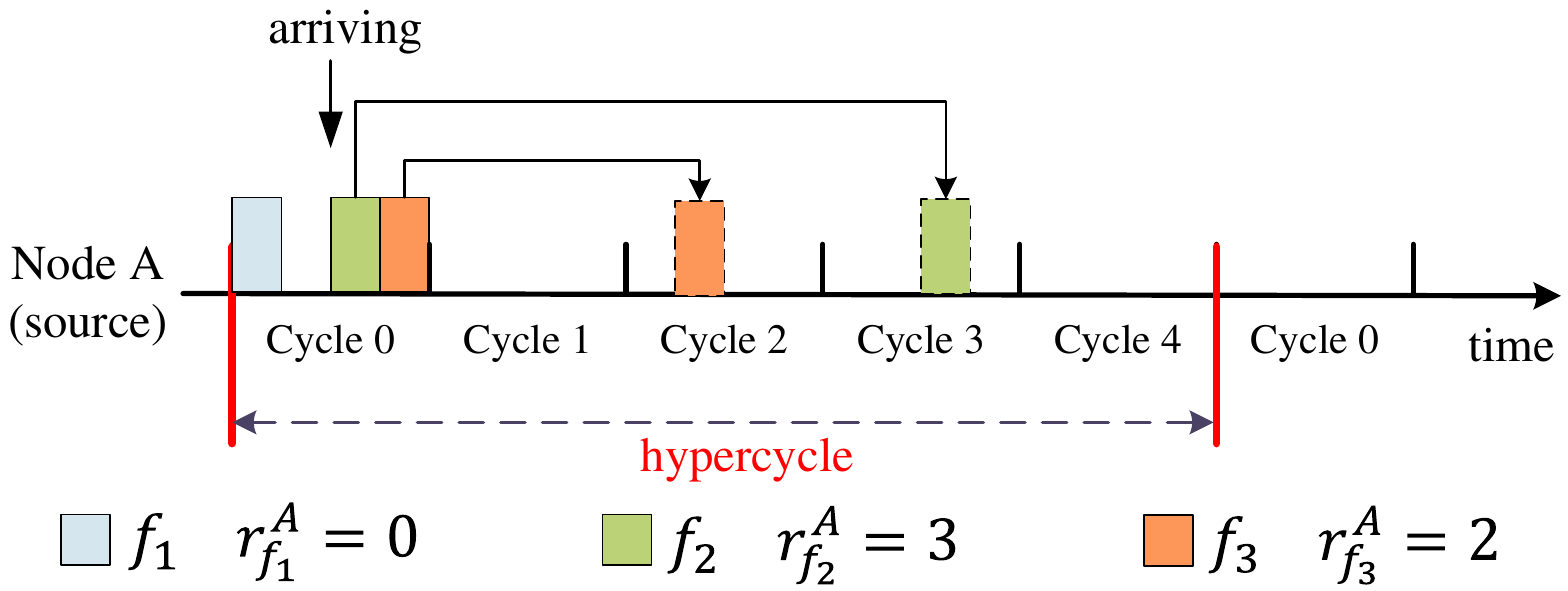}}
\caption{Cycle shifting in sources. When too many TS flows ($f_1$, $f_2$, and $f_3$) arrive in a cycle (cycle 0), cycle shifts are introduced for traffic shaping. $r_{f_1}^A = 0$ represents that $f_1$ will be retransmitted in cycle 0. $r_{f_2}^A = 3$ indicates the retransmitting cycle of $f_2$ is $(0+3) \mod 5 = 3$. $f_3$ will be retransmitted in cycle $\left((0+2) \mod 5 = 2\right)$.    }
\label{fig::cycShiftingSrc}
\end{figure}

{\bf{Cycle shifting in sources}:} 
A cycle shift $r_{f}^A$ is corresponding to the TS flow $f$ and the source $A$. If packets of $f$ arrive at node $A$ in cycle $a$, and a hypercycle contains $N$ cycles, these packets will be retransmitted in cycle $\left((a+r_{f}^A) \mod N\right)$, where $r_{f}^A \in [0,N-1]$. In Fig.~\ref{fig::cycShiftingSrc}, packets of TS flows $f_{2}$ all arrive at cycle 0 from upper layer. Due to $r_{f_2}^A = 3$, $f_{2}$ will be sent out in cycle $\left((0+3) \mod 5\right)$.

{\bf{Cycle shifting in CQF switches/DIP routers}:} In the device $v \in V_{\rm{cqf}} \cup V_{\rm{dip}}$, the cycle shift is always 1. Assume that there is a pair of intermediate nodes $(A,B)$ with $\Phi_{(A,B)}(a) = b$. TS flow $f$ is sent in node $A$'s cycle $a$. A hypercycle contains $N$ cycles in node $B$. Because $r_f^B = 1$, flow $f$ is retransmitted in cycle $(b+1) \mod N$.

\begin{figure}[htbp]
\centerline{\includegraphics[width=.75\linewidth]{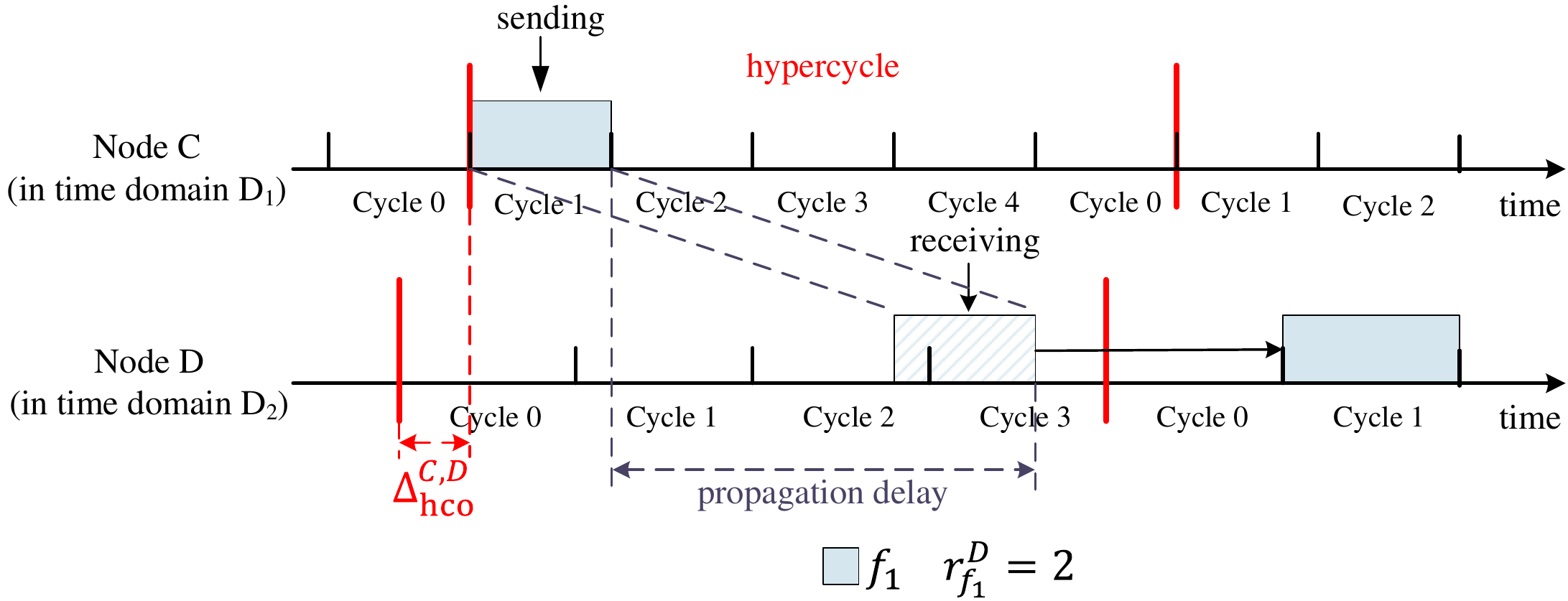}}
\caption{Cycle shifting in edge switches or routers. TS flow $f_1$ is sent in cycle 1 in node $C$, and cycle 0 is mapped to cycle 3 in node $D$ (i.e., $\Phi_{(C,D)}(0) = 3$). A hypercycle contains 4 cycles in node $D$. Since $r_{f_1}^D = 2$, $f_1$ will be retransmitted in cycle $(3+2) \mod 4 = 1$ in node $D$.  }
\label{fig::cycShiftingEdge}
\end{figure}

{\bf{Cycle shifting in edge switches/routers}:} 
The cycle shifting in edge devices is illustrated in Fig.~\ref{fig::cycShiftingEdge}. Node $C$ and node $D$ are in different time domains. A hypercycle contains $N=4$ cycles of node $D$. There exists $\Phi_{(C,D)}(1)=3$. The cycle shift of TS flow $f_{1}$ in node $D$ is $r_{f_{1}}^{D}=2$, so packets of $f_{1}$ will be retransmitted in cycle $(\Phi_{(C,D)}(1)+r_{f_1}^{D})\mod N=1$. The range of $r_{f_1}^D$ is $[0,N-1]$.

\section{Joint scheduling}\label{section4}

\subsection{Decision variables}
There are a series of decision variables to schedule TS flows in $F$. These variables can be classified into three types: admission control, path selection, and cycle shifting.

{\bf{Admission control}:} For a TS flow $f \in F$, $x^f$ is defined to describe whether $f$ should be accepted ($x^f=1$) or not ($x^f=0$). Let ${\textbf{x}} =\{x^f | f \in F\}$.

{\bf{Path selection}:} $p^f = (v_0, v_1, \cdots, v_{|p^f|})$ is defined as the scheduled path for TS flow $f$. $v_0$ is the source and $v_{|p^f|}$ is the destination. Let $P = \{p^f | f \in F\}$.

{\bf{Cycle shifting}:} The cycle shifting for TS flow $f$ is defined as ${\bf r}_{p^f} = (r_f^{v_0}, r_f^{v_1}, \cdots, r_f^{v_{|p^f|}})$, where $r_f^{v_i}$ is the cycle shifts for $f$ at node $v_i$. Let $\textbf{r} = \{{\bf r}_{p^f} | p^f \in P\}$.

\subsection{Constraints}
There exist two constraints in joint scheduling: constraints on end-to-end delays and resource capacity.

{\bf{Constraints on end-to-end delay}:} The end-to-end delay of a TS flow $f$ depends on the scheduled path $p^f$ and the cycle shifting ${\bf r}_{p^f}$. Thus, the end-to-end delay of $f$ is described as $\Delta (p^f, {\bf r}_{p^f})$, and satisfies $\Delta (p^f, {\bf r}_{p_f}) \leq \Delta_f^{\rm{e2e}}$ ($\Delta_f^{\rm{e2e}}$ is the maximum acceptable end-to-end delay of $f$).

For a link $e_i = (v_i, v_{i+1})$ in path $p^f$, we define the delay on $e_i$ as $\Delta_{e_i}^f$, which contains the maximum time staying at $v_i$ and the propagation delay on $e_i$ (i.e., $\Delta_{e_i}$). Thus, the upper bound of $\Delta_{e_i}^f$ can be defined as
\begin{equation}
    U(\Delta_{e_i}^f) = (r_f^{v_i}+1)\Delta^{v_i} + \Delta_{e_i}
\end{equation}
where $\Delta^{v_i}$ is the duration of a cycle in node $v_i$.

The upper bound of path $p^f$'s end-to-end delay can be calculated by

\begin{equation}
    U(\Delta (p^f,{\bf{r}}_{p^f})) = \sum_{i=0}^{|p^f|-1} U(\Delta_{e_i}^f)
\end{equation}

For every TS flow $f \in F$, the constraint of end-to-end delay can be described as

\begin{equation}\label{E2EdelayConstraint}
    U(\Delta (p^f,{\bf{r}}_{p^f})) \leq \Delta_f^{\rm{e2e}}
\end{equation}

{\bf{Constraints on resource capacity}:} The resource in the hierarchical network is cycles in every node. For a cycle $c$ in node $v_i$ on link $e=(v_i, v_{i+1})$, if a set of TS flows $F_c$ is assigned to it, the total bits of $F_c$ should not exceed the transmission capacity of $c$ (i.e., formula \eqref{cycleConstraint}).
\begin{equation}\label{cycleConstraint}
    \sum_{f \in F_c} \omega_f^c \leq \Delta^{v_i} \cdot BW_e
\end{equation}
where $\omega_f^c$ is the total bits of $f$ transmitted in cycle $c$, $\Delta^{v_i}$ is the duration of a cycle in node $v_i$, and $BW_e$ is the bandwidth of link $e$.

\subsection{Objective function}
The target of the joint scheduling is to accept as many $f \in F$ as possible. The priority of every TS flow may be different. A weight value $v^f \in (0,1]$ represents the priority of TS flow $f$. The problem can be formulated as an integer programming (IP):

\begin{subequations}\label{P1}
\begin{alignat}{2}
& \max\limits_{{\bf x}, {P}, {\bf{r}}} \sum_{f \in F} v^f x^f &  \\
& s.t. \quad v^f \in (0,1], x^f \in \{0,1\}&  \\
& \quad\quad\quad \eqref{E2EdelayConstraint},\eqref{cycleConstraint}& 
\end{alignat}
\end{subequations}

\section{Simulation}\label{section5}

The hierarchical network shown in Fig.~\ref{fig::structureOfConvergedNetworks} is constructed in the simulation. The topology of the core network containing 15 DIP  routers (with 10~$\mu$s as $\Delta_{\rm{dip}}$) is established based on the network model Atlanta \cite{OrlowskiPioroTomaszewskiWessaely2010}. The lengths of links in the core network are 30~km to simulate long-distance transmission. The bandwidth of ports on DIP routers is 10~Gbps. We deploy 10 access networks, and every access network contains 2 CQF switches (with 25~$\mu$s as $\Delta_{\rm{cqf}}$). The bandwidth of ports on CQF switches is 1~Gbps \cite{yan2020injection}. TS flows are connected to CQF switches. Packets of TS flows have to pass through the access network containing the source hosts, the core network, and the access network connected by the destination hosts.

Each access network contains 200 TS flows. The data rate of each TS flow is 500~kbps. Thus, the size of packets is 500~bit, and the interval of packets is 1ms. Besides, the maximum acceptable end-to-end delay of these TS flows is 1~ms. The schedule of packets is conducted in a greedy algorithm. 
In the simulation, the best-effort interference flows with a packet size of 1500~Byte are injected into the converged network. For verifying the effectiveness of the proposed joint scheduling in this paper, two experiments are conducted: {\bf{(1)}} In different levels of interference flows, we compare the end-to-end delays of jointly scheduled flows and best-effort transmitted flows to verify the capability to providing bounded delays and jitters; {\bf{(2)}} To prove the effectiveness of traffic shaping and path selection, the comparison of flows admission in three strategies (i.e., with traffic shaping and path selection, without traffic shaping, and without path selection) is performed.

\begin{figure}[htbp]
\centerline{\includegraphics[width=.65\linewidth]{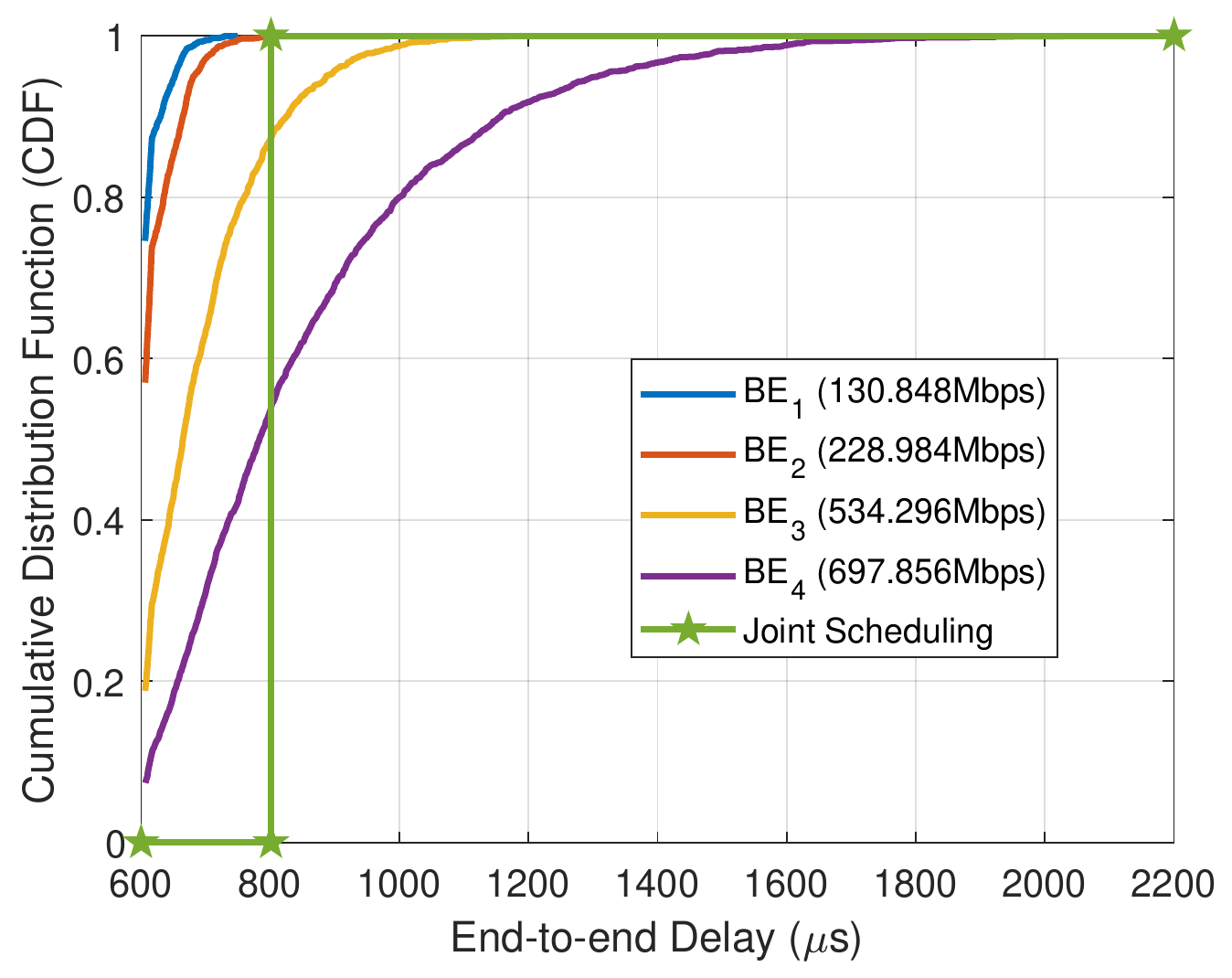}}
\caption{The cumulative distribution of end-to-end delays in different intensities of interference flows. With the increase of interference flows intensity, the delays and jitters of the TS flow transported in best-effort (BE) increase sharply. However, the joint scheduling can provide a constant end-to-end delay in 801~$\mu$s, and zero jitters. }
\label{fig::boundedDelayJitter}
\end{figure}

Tabel \ref{tab::delaysComp} and Fig.~\ref{fig::boundedDelayJitter} show the result of experiment 1, and prove the effectiveness of guaranteeing bounded end-to-end delays and jitters. We randomly choose a TS flow in an access network to observe. The TS flow is transmitted in best-effort and the proposed joint scheduling. The total throughput of all 200 TS flows in the access network is 100~Mbps. With the increase of intensity (i.e., throughput) of interference flows, the end-to-end delays using best-effort (BE) grow significantly. In the light-load scenario injecting interference flows with throughput 229~Mbps, the maximum delay is 980~$\mu$s, and the jitter is 375~$\mu$s. In the medium-load scenario with 534~Mbps interference flows, the average delays are higher compared with light-load scenarios. The maximum delay is up to 1173~$\mu$s, and the jitter is up to 567~$\mu$s. Besides, there exist 26 TS flows having end-to-end delays beyond the maximum acceptable delay (1~ms). The delays using the joint scheduling is a constant 801~$\mu$s with zero jitters. Delays are not affected by the intensity of interference flows. In an extremely light-load scenario (i.e., the throughput of interference flows is 131Mbps), the maximum delay of best-effort transmission is less than the one in deterministic transmission. This phenomenon results from the extra delays introduced from cyclic transmission fashion in deterministic networking and cycle shifting. For guaranteeing bounded end-to-end delays and jitters, the extra delays are acceptable, and the delays in deterministic transmission are significantly less than the maximum delay in light-load, medium-load, and high-load scenarios.

\begin{table*}[]
\caption{The statistics of end-to-end delays in BE and joint scheduling}\label{tab::delaysComp}
\begin{tabular}{|c|c|c|c|c|c|c|c|c|}
\hline
\multirow{2}{*}{\begin{tabular}[c]{@{}c@{}}Intensity of interference flows \\ (Mbps)\end{tabular}} & \multicolumn{4}{c|}{Best-Effort}                                                                                                      & \multicolumn{4}{c|}{Joint scheduling}                                                                                                 \\ \cline{2-9} 
                                                                                                   & \multicolumn{1}{l|}{mean($\mu$s)} & \multicolumn{1}{l|}{max($\mu$s)} & \multicolumn{1}{l|}{jitter($\mu$s)} & \multicolumn{1}{l|}{beyond deadline} & \multicolumn{1}{l|}{mean($\mu$s)} & \multicolumn{1}{l|}{max($\mu$s)} & \multicolumn{1}{l|}{jitter($\mu$s)} & \multicolumn{1}{l|}{beyond deadline} \\ \hline
130.848                                                                                            & 612                           & 750                          & 144                             & 0                                    & 801                           & 801                          & 0                               & 0                                    \\ \hline
228.984                                                                                            & 621                           & 980                          & 375                             & 0                                    & 801                           & 801                          & 0                               & 0                                    \\ \hline
534.296                                                                                            & 692                           & 1173                         & 567                             & 26                                   & 801                           & 801                          & 0                               & 0                                    \\ \hline
697.856                                                                                            & 847                           & 2031                         & 1426                            & 402                                  & 801                           & 801                          & 0                               & 0                                    \\ \hline
\end{tabular}
\end{table*}

\begin{figure}[htbp]
\centerline{\includegraphics[width=.65\linewidth]{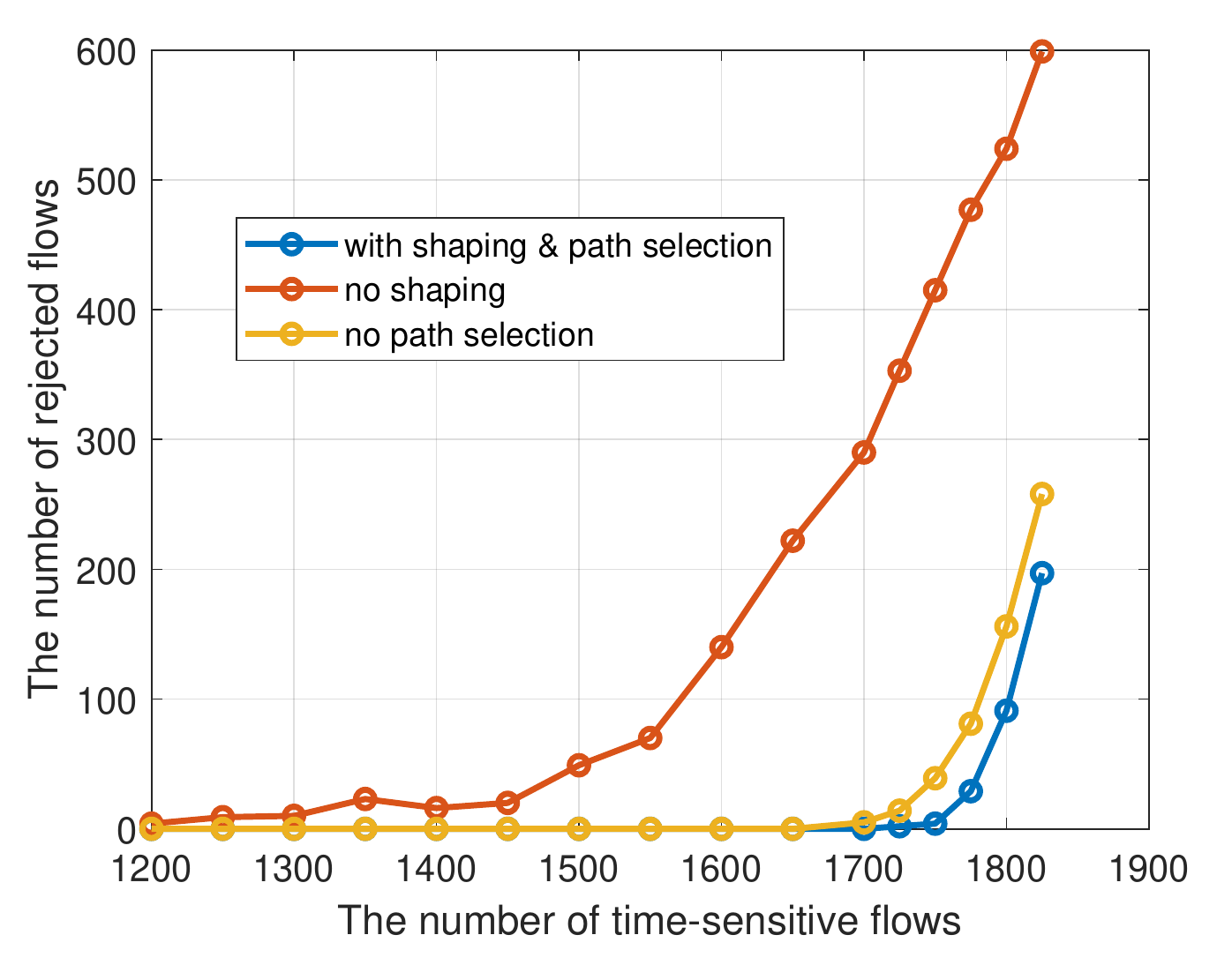}}
\caption{The number of rejected TS flows in different intensities of TS flows. Compared with scheduling without traffic shaping or path selection, the proposed joint scheduling can decrease the number of rejected TS flows. Especially compared with scheduling without traffic shaping, the proposed scheduling can increase network throughput.  }
\label{fig::flowAdmission}
\end{figure}

Fig.~\ref{fig::flowAdmission} illustrates the result of experiment 2. For verifying the proposed joint scheduling can raise the throughput of TS flows, we deploy a different number of TS flows that are waiting for scheduling in every access network. Obviously, the total throughput of these TS flows is less than the capacity of links. Using the joint scheduling with traffic shaping and path selection, the number of TS flows rejected for deterministic transmission starts increasing when every access network contains 1725 TS flows. Without path selection (i.e., all flows choose the path with a minimum number of hops), the number of rejected flows starts increasing when 1700 TS flows wait for scheduling in an access network. Moreover, with the same number of TS flows, the number of rejected flows is higher than the one using joint scheduling. Without traffic shaping, the scheduling starts to reject TS flows when there exist 1200 TS flows. The number of rejected flows starts to increase sharply when an access network contains 1500 TS flows. The rejection is severer than the joint scheduling and the scheduling without path selection. The result shows the traffic shaping and path selection can improve the throughput of TS flows, and the traffic shaping is the more important factor to improve the throughput. Due to traffic shaping can deploy flows to different transmission cycles, the conflict for transmission resources is reduced. However, because of the tree topology in access networks, all flows in the same access network have to pass through common CQF switches, which attenuate the influence of path selection. The increase of these three curves is not linear, because the newly added TS flows may have a high priority to be scheduled preferentially, and this will make the TS flows accepted formerly being rejected. Thus, the increase is not linear.

\section{Conclusion}\label{section6}
To empower the end-to-end deterministic transmission, this paper proposes a hierarchical network that consists of the access networks and the core networks.
In the access network, we use the CQF technique to ensure the deterministic transmission during the aggregation of the traffic from hosts.
In the core network, DIP is exploited for the long-distance deterministic transmission over the backbone links. The end-to-end deterministic transmissions are realized by the cycle alignment and traffic shaping at the network edge. Moreover, a joint scheduling approach is also formulated to improve the network throughput. Simulations results based on the real-world networks from sndlib show that the proposed network can achieve a deterministic transmission even in the high-load scenarios.


\section*{Acknowledgment}
This work was supported by National Key Research and Development Program of China (Grant No. 2020YFB1805200, No. 2020YFB1806400).



%
\vspace{12pt}

\end{document}